\documentclass[epsf,prb,twocolumn,showpacs,nofootinbib]{revtex4-1}

\usepackage[pdftex]{graphicx}
\usepackage{dcolumn}
\usepackage{bm}
\usepackage{epsfig}
\usepackage{latexsym}
\usepackage{amsmath}
\usepackage{amsfonts}
\usepackage{amssymb}
\usepackage{color}
\usepackage{array}
\usepackage{framed}

\begin{document}

\title{Higher-Spin Witten Effect and \\Two-Dimensional Fracton Phases}
\author{Michael Pretko \\
\emph{Center for Theory of Quantum Matter, University of Colorado, Boulder, CO 80309}}
\date{September 7, 2017}

\begin{abstract}
We study the role of ``$\theta$ terms" in the action for three-dimensional $U(1)$ symmetric tensor gauge theories, describing quantum phases of matter hosting gapless higher-spin gauge modes and gapped subdimensional particle excitations, such as fractons.  In Maxwell theory, the $\theta$ term is a total derivative which has no effect on the gapless photon, but has two important, closely related consequences: attaching electric charge to magnetic monopoles (the Witten effect) and leading to a Chern-Simons theory on the boundary.  We will find that a similar story holds in the higher-spin $U(1)$ gauge theories.  These theories admit generalized $\theta$ terms which have no effect on the gapless gauge mode, but which bind together electric and magnetic charges (both of which are generally subdimensional) in specific combinations, in a higher-spin manifestation of the Witten effect.  We derive the corresponding Witten quantization condition.  We find that, as in Maxwell theory, imposing time-reversal invariance restricts $\theta$ to certain discrete values.  We also find that these new $\theta$ terms imply a non-trivial boundary structure.  The boundaries host fracton excitations coupled to a tensor $U(1)$ gauge field with a Chern-Simons-like action, in both chiral and non-chiral varieties.  These boundary theories open a door to the study of $U(1)$ fracton phases described by tensor Chern-Simons theories, not only on boundaries of three-dimensional systems, but also in strictly two spatial dimensions.  We explicitly work through three examples of bulk and boundary theories, the principles of which can be readily extended to arbitrary higher-spin theories.
\end{abstract}

\maketitle

\tableofcontents

\section{Introduction}

Quantum phases of matter, such as spin liquids and quantum hall systems, are well-described in the language of gauge theory.  While most studies have focused on familiar vector gauge theories, recent theoretical efforts have established the existence of stable three-dimensional phases described by higher rank symmetric tensor gauge fields.  These include both $U(1)$ tensor gauge theories with gapless higher-spin gauge excitations\cite{alex,sub,genem}, and the ``generalized lattice gauge theories" of Vijay, Haah, and Fu, which are the natural discrete analogue.\cite{fracton1,fracton2}  These tensor gauge theories are particularly interesting due to the fact that the gauge charges necessarily behave as subdimensional particles - excitations which are restricted by gauge invariance to move only within lower-dimensional subspaces of the three-dimensional system.  As examples, particles can be restricted to motion within a plane, a line, or even a single point.  In this last case, where the excitation is completely immobile, the particle is known as a ``fracton."  These exotic new particles have been a topic of intense recent research\cite{sub,genem,fracton1,fracton2,chamon,bravyi,cast,yoshida,haah,haah2,williamson,sagarlayer,hanlayer,abhinav,mach,parton,slagle,bowen,nonabel,balents}, since they sit at the intersection of numerous areas of modern theoretical physics, including long-range entanglement\cite{fracton1,fracton2,bowen}, quantum error correction\cite{haah,haah2,nonabel}, glassy dynamics\cite{chamon,abhinav,screening}, and emergent gravity.\cite{mach}

In this paper, we will focus on the $U(1)$ higher-spin theories, which are in some sense the simplest since they build directly on intuition from the Maxwell theory of a conventional $U(1)$ vector gauge field.  We therefore expect many of the interesting stories associated with Maxwell theory to have natural tensor analogues.  In particular, we will here investigate the role of $\theta$ terms in the action for the higher-spin theories.  These terms are natural analogues of the ``$E\cdot B$" term (conventionally parameterized by coefficient $\theta$) which can be present in the action for Maxwell theory.  The conventional $\theta$ term has no effect on the behavior of the gapless photon.  However, it causes the fundamental magnetic monopole of the system to pick up an electric charge given by $\theta/2\pi$ times the fundamental electric charge.  This changes the set of dyons (particles carrying both electric and magnetic charge) which can occur in the theory, via the Witten quantization condition.\cite{witten}  This shift in the dyon structure is often referred to as the ``Witten effect," which has received renewed attention in recent years due to its role in topological insulators and other symmetry protected topological phases.\cite{qi,rosen,spt}  We will find that a similar story holds in the higher-spin $U(1)$ gauge theories.  The action for each of the higher-spin theories allows for a natural generalization of the $\theta$ term, with similar properties.  As before, the gapless gauge mode will be untouched, but the magnetic charges will pick up some amount of electric charge, modifying the dyon structure of the theory.  We will generalize Witten's quantization condition, relating the $\theta$ parameter to the amount of electric charge picked up by the fundamental monopole.  The details of the charge attachment will differ depending on the precise structure of the $\theta$ term, and we will find that some systems admit multiple different types of $\theta$ terms.  We explicitly work through three rank 2 examples to illustrate the general principle, which can then be readily extended to theories of arbitrary rank.

Another famous feature of the standard $\theta$ term is its relationship with Chern-Simons theory.  The $\theta$ term is actually a total derivative term in the action, which is why it has no effect on the low-energy gauge-mode.  Nevertheless, this total derivative leads to a Chern-Simons term living on the boundary which will attach electric charge to a monopole passing into the system, causing the Witten effect.\cite{spt,chong}  This teaches us that quantum hall phases (described by Chern-Simons theory) can naturally occur on the boundary of certain bulk systems coupled to Maxwell theory, such as topological insulators.  And for systems with an emergent $U(1)$ gauge field, such as $U(1)$ spin liquids, the edge can naturally host topologically ordered phases.  The $\theta$ term therefore is not only useful for studying the bulk physics of Maxwell theory.  It also provides a novel way of studying the physics of two-dimensional quantum hall systems.  In close analogy, we will find that the higher-spin $\theta$ terms also lead to interesting two-dimensional phases on the boundary.  In this case, however, these will not be phases which we already independently understand.  Rather, the higher-spin $\theta$ terms open the door to a set of new two-dimensional phases hosting fracton excitations, via tensor generalizations of Chern-Simons theory.  These will include both chiral and non-chiral varieties.  While we first encounter these phases at the boundary of three-dimensional systems, they can also occur in strictly two-dimensional systems under appropriate conditions, as we will discuss.  This will be the first time the fracton phenomenon has been stably manifested in two dimensions.

\section{Lagrangians}

In order to discuss the conventional $\theta$ term and its generalizations, it is most useful to work in the Lagrangian formulation.  However, for tensor gauge theories, essentially all previous work in the condensed matter literature has taken place in the Hamiltonian language.\cite{alex,sub,genem,fracton1,fracton2,chamon,bravyi,cast,yoshida,haah,haah2,williamson,sagarlayer,hanlayer,abhinav,mach,parton,slagle,bowen,nonabel,balents}  For gauge theories of this sort, going between the two formulations can be a nontrivial task.  We will therefore begin by discussing how to correctly formulate Lagrangians for higher-spin $U(1)$ gauge theories.

\subsection{Review of Maxwell Theory}

We begin by reviewing how to go between the Lagrangian and Hamiltonian formulations of standard three-dimensional Maxwell theory.  We start from the Lagrangian formulation, since this is perhaps more familiar in the field-theoretic context, and use it to derive the Hamiltonian.  We can write the partition function of the theory in terms of the standard Maxwell action:
\begin{equation}
Z = \int \mathcal{D}A_\mu e^{i\int d^4x \mathcal{L}} = \int \mathcal{D}A_\mu\, e^{i\int d^4x\frac{1}{4} F^{\mu\nu}F_{\mu\nu}}
\label{plain}
\end{equation}
where the field strength is $F_{\mu\nu} = \partial_\mu A_\nu - \partial_\nu A_\mu$.  (For readers concerned about issues of gauge fixing, this will be discussed in Appendix C.)  More explicitly, in terms of the spatial vector potential $\vec{A}$ and the timelike component $A_0$, we can write this as:
\begin{align}
\begin{split}
Z = \int \mathcal{D}\vec{A}\mathcal{D}A_0&\exp\bigg[i\int d^3x dt \bigg(\\
&\frac{1}{2}(\dot{\vec{A}}-\vec{\nabla}A_0)^2 - \frac{1}{2} (\vec{\nabla}\times \vec{A})^2\bigg)\bigg]
\end{split}
\end{align}
where dots indicate time derivatives.  To convert to the Hamiltonian formulation, we introduce an auxiliary field $\vec{E}$, which will eventually play the role of the electric field in the Hamiltonian.  Inserting an appropriate Gaussian integral over the field $\vec{E}$, we can write the path integral as:
\begin{align}
\begin{split}
Z \propto \int \mathcal{D}\vec{A}&\mathcal{D}A_0\mathcal{D}\vec{E} \exp\bigg[i\int d^3xdt\bigg(\\
&\vec{E}\cdot (\dot{\vec{A}}-\vec{\nabla}A_0) -\frac{1}{2}E^2 - \frac{1}{2}(\vec{\nabla}\times\vec{A})^2\bigg)\bigg]
\end{split}
\end{align}
(where we ignore overall multiplicative constants in the partition function).  The exponent is now linear in the timelike component $A_0$, so after an integration by parts, we can integrate $A_0$ out of the theory entirely:
\begin{align}
\begin{split}
Z \propto \int \mathcal{D}\vec{A}\mathcal{D}\vec{E} \,\,\delta&(\vec{\nabla}\cdot \vec{E}) \,\exp\bigg[i\int d^3xdt\bigg(\\
&\vec{E}\cdot \dot{\vec{A}}-\frac{1}{2}E^2 - \frac{1}{2}(\vec{\nabla}\times\vec{A})^2\bigg)\bigg]
\label{zham}
\end{split}
\end{align}
We now have a path integral which enforces the rigid constraint that $\vec{\nabla}\cdot \vec{E} = 0$ everywhere.  (Note that the delta function above actually represents a product of delta functions at each point in space.)  We can recognize this as the appropriate Hamiltonian formulation of the path integral.  For comparison, a single-particle path integral can be written in equivalent Lagrangian and Hamiltonian forms as:
\begin{equation}
Z_{s.p.} \propto \int \mathcal{D}x \,e^{i\int dt L}\propto \int \mathcal{D}x\mathcal{D}p\, e^{i\int dt (p\dot{x} - H)}
\end{equation}
Along the same lines, we can use Equation \ref{zham} to read off the Hamiltonian density of our theory as:
\begin{equation}
\mathcal{H} = \frac{1}{2}E^2 + \frac{1}{2}(\vec{\nabla}\times\vec{A})^2
\label{ham}
\end{equation}
where $\vec{A}$ and $\vec{E}$ are canonical conjugate variables, and the system is subject to the constraint:
\begin{equation}
\vec{\nabla}\cdot \vec{E} = 0
\end{equation}
We see that we recover precisely the expected $\frac{1}{2}(E^2+B^2)$ behavior of the classical electromagnetic Hamiltonian, along with a constraint expressing the absence of charge in the pure gauge theory.

This is exactly the sort of Hamiltonian structure that arises in systems described by emergent $U(1)$ gauge fields, such as $U(1)$ spin liquids.  The constraint arises from some local energetic considerations, such as a spin-ice rule.  The $E^2$ and $B^2$ terms enter as the most relevant gauge-invariant terms.  The constraint structure of the microscopic Hamiltonian thereby leads directly to the Maxwell Hamiltonian as the effective low-energy description of the spin liquid phase.  One could then perform the entire analysis above in reverse, demonstrating that the Maxwell action is the appropriate low-energy description of the $U(1)$ spin liquid.

Of course, at higher energies, there will also be states which violate energetic constraints like a spin-ice rule.  This corresponds to states with $\vec{\nabla}\cdot\vec{E} = \rho \neq 0$, or in other words, states with charges.  For the purposes of this paper, we will not need to know the detailed dynamics of these charges.  All that we need to know is that these charges exist, and also are quantized: $\vec{\nabla}\cdot\vec{E} = ne$, for integer $n$ and fundamental charge $e$.  This quantization is a result of the fact that the gauge field can generically be compact, $A_i\sim A_i +R$, for some compactification radius $R$ satisfying $eR = 2\pi$.  For convenience, we will take $e = 1$ and $R = 2\pi$ throughout.  These charges will play an important role when we come to discuss the Witten effect.

\subsection{Higher-Spin Theories}

We now turn our attention towards the higher-spin $U(1)$ gauge theories, which host particles with subdimensional behavior, such as fractons.  From the Hamiltonian formulations described in earlier work\cite{alex,sub,genem}, we will show how to transition to a Lagrangian formulation.  We will work through the details for two specific phases described by rank 2 symmetric tensor gauge fields to demonstrate the general principles.  By working with tensors of the appropriate rank and constraints, these principles can be extended to any other higher-spin theory.

\subsubsection{Vector Charge Theory}

We consider a phase described by a rank 2 spatial symmetric tensor gauge field $A_{ij}$, with a canonical conjugate variable $E_{ij}$, playing the role of an electric field tensor.  As described in previous work\cite{alex,sub}, there are multiple different stable theories with the same degrees of freedom.  Each theory can be uniquely specified by its gauge constraint, corresponding to a generalized Gauss's law.  In the first theory that we will consider, this Gauss's law takes the form:
\begin{equation}
\partial_i E^{ij} = \rho^j
\end{equation}
for vector charge density $\rho^j$.  (All indices are spatial, and repeated indices are summed over.)  The particles carrying this vector charge are notable for obeying an extra conservation law which has no analogue in a conventional gauge theory.  These charges obey the following two conservation laws:
\begin{equation}
\int\vec{\rho} \,=\, \textrm{constant}\,\,\,\,\,\,\,\,\,\,\,\int\vec{x}\times\vec{\rho} \,=\, \textrm{constant}
\end{equation}
where $\vec{x}$ is the spatial coordinate and the integrals run over three-dimensional space.  The first equation is the natural analogue of charge conservation.  The second tells us that the angular moment associated with this charge vector is an independently conserved quantity.  This second conservation law has a fairly drastic consequence for the charges: an isolated charge can only move in the direction of its charge vector, while transverse motion is strictly forbidden.

In the low-energy sector, without any charges, we have the more restrictive form of Gauss's law, $\partial_iE^{ij} = 0$, which leads to invariance under the following gauge transformation:
\begin{equation}
A_{ij}\rightarrow A_{ij} + \partial_i\alpha_j + \partial_j\alpha_i
\end{equation}
where the vector gauge parameter $\alpha_i$ has arbitrary spatial dependence.  Including the most relevant terms consistent with the gauge constraint, the low-energy Hamiltonian for this phase is given by:
\begin{equation}
\mathcal{H} = \frac{1}{2}E^{ij}E_{ij} + \frac{1}{2}B^{ij}B_{ij}
\end{equation}
where the magnetic tensor takes the form:
\begin{equation}
B_{ij} = \epsilon_{iab}\epsilon_{jcd}\partial^a\partial^c A^{bd}
\label{magtensor}
\end{equation}
and we implicitly impose the constraint $\partial_iE^{ij} = 0$.  This Hamiltonian leads to gapless gauge modes, including a spin-2 mode with quadratic dispersion.  (There are also lower-spin gauge modes, which could be eliminated by more complicated gauge constraints, if desired.)

We can now take this constrained Hamiltonian structure and use it to reverse all of the logic from the previous section, going to a Lagrangian formulation.  From the Hamiltonian, we can write down the partition function as:
\begin{align}
\begin{split}
Z \propto \int \mathcal{D}A_{ij}\mathcal{D}&E_{ij} \,\delta(\partial_iE^{ij})\exp\bigg[i\int\bigg(\\
&E^{ij}\dot{A}_{ij} - \frac{1}{2}E^{ij}E_{ij} - \frac{1}{2}B^{ij}B_{ij}\bigg)\bigg]
\end{split}
\end{align}
where the integral in the exponent is over space and time (measure suppressed for convenience), and the delta function enforces that each component of $\partial_iE^{ij}$ vanishes at each point in space.  In the case of Maxwell theory, the gauge constraint was enforced by a Lagrange multiplier, which happened to play the role of a timelike component to the gauge field, $A_0$.  In the present case, we can also introduce a Lagrange multiplier field to handle the constraint.  However, this higher-spin theory does not possess any relativistic symmetry (the dispersion is quadratic), and there is no reason to expect the Lagrange multipliers to behave as timelike components to $A_{ij}$.  We therefore abandon relativistic notation and write our Lagrange multiplier field as an independent vector variable, $C_i$.  We can then write our partition function as:
\begin{align}
\begin{split}
Z\propto &\int\mathcal{D}A_{ij}\mathcal{D}E_{ij}\mathcal{D}C_i \exp\bigg[i\int\bigg(\\
&C_j\partial_i E^{ij} + E^{ij}\dot{A}_{ij} - \frac{1}{2}E^{ij}E_{ij} - \frac{1}{2}B^{ij}B_{ij}\bigg)\bigg]
\end{split}
\end{align}
Integrating the $C_i$ term by parts and taking advantage of the symmetry of $E^{ij}$, we can write this as:
\begin{align}
\begin{split}
Z\propto \int&\mathcal{D}A_{ij}\mathcal{D}E_{ij}\mathcal{D}C_i \exp\bigg[i\int\bigg(\\
E^{ij}&(\dot{A}_{ij} - \frac{1}{2}\partial_iC_j - \frac{1}{2}\partial_jC_i) - \frac{1}{2}E^{ij}E_{ij} - \frac{1}{2}B^{ij}B_{ij}\bigg)\bigg]
\end{split}
\end{align}
The field $E^{ij}$ can now be integrated out of the problem to yield:
\begin{align}
\begin{split}
Z\propto \int &\mathcal{D}A_{ij}\mathcal{D}C_i \exp\bigg[i\int\bigg(\\
&\frac{1}{2}(\dot{A}_{ij}-\frac{1}{2}\partial_iC_j - \frac{1}{2}\partial_jC_i)^2 - \frac{1}{2}B^{ij}B_{ij}\bigg)\bigg]
\end{split}
\end{align}
We can then read off the Lagrangian of our theory as:
\begin{equation}
\mathcal{L}[A_{ij},C_i] = \frac{1}{2}(\dot{A}_{ij}-\frac{1}{2}\partial_iC_j - \frac{1}{2}\partial_jC_i)^2 - \frac{1}{2}B^{ij}B_{ij}
\label{veclag}
\end{equation}
We now have a Lagrangian description of a theory with a tensor field coupled to a vector field, with a larger set of gauge transformations than our original Hamiltonian formulation.  While we originally spoke only in terms of spatial gauge transformations, we can now identify the full time-dependent gauge transformations of our theory as:
\begin{equation}
A_{ij}\rightarrow A_{ij}+\partial_i\alpha_j +\partial_j\alpha_i
\end{equation}
\begin{equation}
C_i\rightarrow C_i + \dot{\alpha}_j
\end{equation}
where the vector gauge parameter $\alpha(x,t)$ now has arbitrary dependence on both space and time.

\subsubsection{Scalar Charge Theory}

We now switch to a different rank 2 theory, which is also described by a symmetric tensor gauge field $A_{ij}$ with canonical conjugate $E_{ij}$, but with a different Gauss's law, given by:
\begin{equation}
\partial_i\partial_j E^{ij} = \rho
\end{equation}
for scalar charge $\rho$.  As in the previous theory, this Gauss's law leads to two separate conservation laws:
\begin{equation}
\int\rho \,=\, \textrm{constant}\,\,\,\,\,\,\,\,\,\,\,\,\int\rho\vec{x}\, =\, \textrm{constant}
\end{equation}
The first equation is simply the conventional conservation of charge, while the second represents the conservation of dipole moment.  This extra conservation law forces an isolated charge to be a fracton excitation, unable to move in any direction, since any motion would change the dipole moment of the system.

In the low-energy sector, which is free of charges, we have the constraint $\partial_i\partial_jE^{ij} = 0$, which leads to invariance under the gauge transformation:
\begin{equation}
A_{ij}\rightarrow A_{ij} + \partial_i\partial_j\alpha
\end{equation}
for gauge parameter $\alpha$ with arbitrary spatial dependence.  The low-energy Hamiltonian consistent with the gauge structure is:
\begin{equation}
\mathcal{H} = \frac{1}{2}E^{ij}E_{ij} + \frac{1}{2}B^{ij}B_{ij}
\end{equation}
where we implicitly impose the constraint $\partial_i\partial_jE^{ij} = 0$, and the magnetic tensor takes the (non-symmetric) form:
\begin{equation}
B_{ij} = \epsilon_{iab}\partial^a A^b_{\,\,\,j}
\end{equation}
(There are other gauge-invariant magnetic tensors which could be formed, such as the one in Equation \ref{magtensor}, but the version above has the fewest number of derivatives and has equal velocities for all of the physical gauge modes, representing a stable fixed point of the renormalization group.)  This Hamiltonian leads to a gapless spin-2 mode with linear dispersion, along with lower-spin gauge modes.

Following the same logic as the previous section, we write down the partition function as:
\begin{align}
\begin{split}
Z \propto \int \mathcal{D}A_{ij}&\mathcal{D}E_{ij} \,\delta(\partial_i\partial_jE^{ij})\exp\bigg[i\int\bigg(\\
&E^{ij}\dot{A}_{ij} - \frac{1}{2}E^{ij}E_{ij} - \frac{1}{2}B^{ij}B_{ij}\bigg)\bigg]
\end{split}
\end{align}
where the delta function enforces the constraint that $\partial_i\partial_jE^{ij}$ vanishes at each point in space.  For this case, we only require one Lagrange multiplier, which we write as a scalar field $\phi$.  We can then write our partition function as:
\begin{align}
\begin{split}
Z\propto \int\mathcal{D}A_{ij}\mathcal{D}E_{ij}\mathcal{D}\phi \exp\bigg[i\int\bigg(&\\
-\phi\partial_i\partial_j E^{ij} + E^{ij}\dot{A}_{ij} - \frac{1}{2}E^{ij}&E_{ij} - \frac{1}{2}B^{ij}B_{ij}\bigg)\bigg]
\end{split}
\end{align}
Integration by parts on the $\phi$ term yields:
\begin{align}
\begin{split}
Z\propto \int\mathcal{D}A_{ij}\mathcal{D}E_{ij}\mathcal{D}\phi \exp\bigg[i\int\bigg(&\\
E^{ij}(\dot{A}_{ij} - \partial_i\partial_j\phi) - \frac{1}{2}E^{ij}E_{ij}& - \frac{1}{2}B^{ij}B_{ij}\bigg)\bigg]
\end{split}
\end{align}
The field $E^{ij}$ can then be integrated out of the problem to yield:
\begin{align}
\begin{split}
Z\propto \int \mathcal{D}&A_{ij}\mathcal{D}\phi \exp\bigg[i\int\bigg(\\
&\frac{1}{2}(\dot{A}_{ij}-\partial_i\partial_j\phi)^2 - \frac{1}{2}B^{ij}B_{ij}\bigg)\bigg]
\end{split}
\end{align}
We read off the Lagrangian of the theory as:
\begin{equation}
\mathcal{L}[A_{ij},\phi] = \frac{1}{2}(\dot{A}_{ij}-\partial_i\partial_j\phi)^2 - \frac{1}{2}B^{ij}B_{ij}
\label{scalag}
\end{equation}
We now have a Lagrangian description of a theory with a tensor field coupled to a scalar field, which once again has a larger set of time-dependent gauge transformations:
\begin{equation}
A_{ij}\rightarrow A_{ij}+\partial_i\partial_j\alpha
\end{equation}
\begin{equation}
\phi\rightarrow\phi + \dot{\alpha}
\end{equation}
where the parameter $\alpha(x,t)$ has arbitrary dependence on both space and time.

\section{$\theta$ Terms and the Witten Effect}

Now that we have Lagrangians in hand, it is time to investigate the possibility of $\theta$ terms.  We will start by giving a simple treatment of the standard $\theta$ term of Maxwell theory, which will help us identify the principles necessary for generalization to the higher-spin theories.

\subsection{Review of Maxwell Theory}

The standard Maxwell action takes the form:
\begin{equation}
S = \int \frac{1}{4}F^{\mu\nu}F_{\mu\nu}
\end{equation}
However, this is not the most general action we can write consistent with gauge invariance.  We may also consider a $\theta$ term (an ``$E\cdot B$" term) of the form:
\begin{align}
S_\theta &= \frac{\theta}{32\pi^2} \int \epsilon^{\mu\nu\rho\sigma}F_{\mu\nu}F_{\rho\sigma} \\
&= \frac{\theta}{4\pi^2}\int (\dot{\vec{A}}-\vec{\nabla}A_0)\cdot \vec{B}
\end{align}
We can immediately conclude that this term will have no effect on the low-energy gauge mode in the bulk of the system, due to the fact that it is the integral of a total derivative:
\begin{equation}
S_\theta = \frac{\theta}{8\pi^2} \int \partial_\mu (\epsilon^{\mu\nu\rho\sigma}A_\nu \partial_\rho A_\sigma)
\end{equation}
The $\theta$ term therefore has no effect on the photon.  It will, however, lead to a Chern-Simons term on a boundary between the system and an external region with $\theta = 0$.  Such a term can attach charge and magnetic flux on the boundary, with the net result that a magnetic monopole brought into the system from the $\theta=0$ region will pick up a specific amount of electric charge.

\begin{figure}[t!]
 \centering
 \includegraphics[scale=0.3]{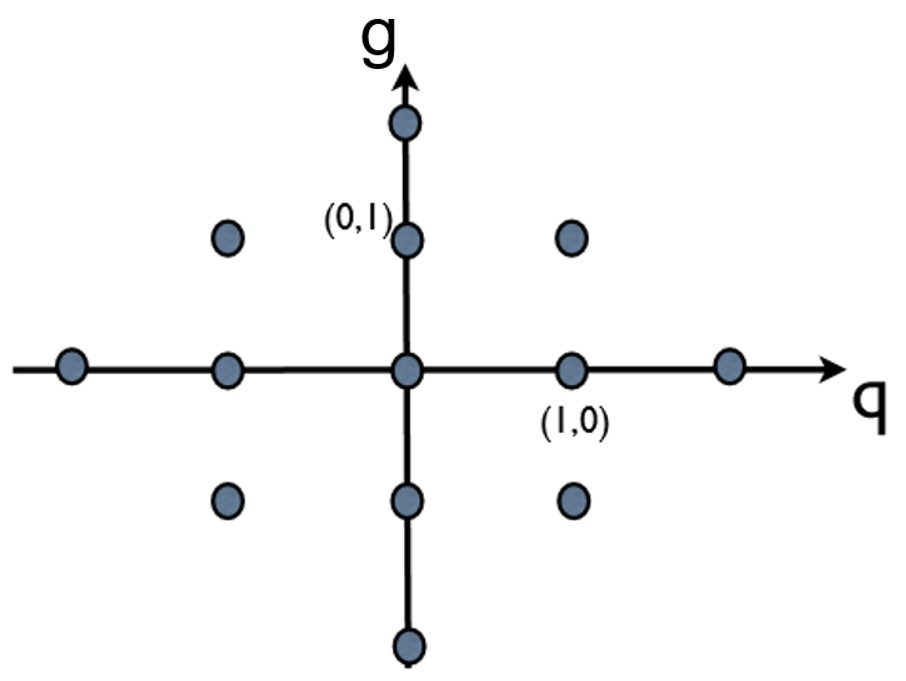}
 \caption{When $\theta=0$, the set of possible charges forms a simple square lattice in the $(q,g)$ plane.  (Adapted from Reference \onlinecite{chong}.)}
 \label{fig:thetazero}
 \end{figure}

\begin{figure}[t!]
 \centering
 \includegraphics[scale=0.34]{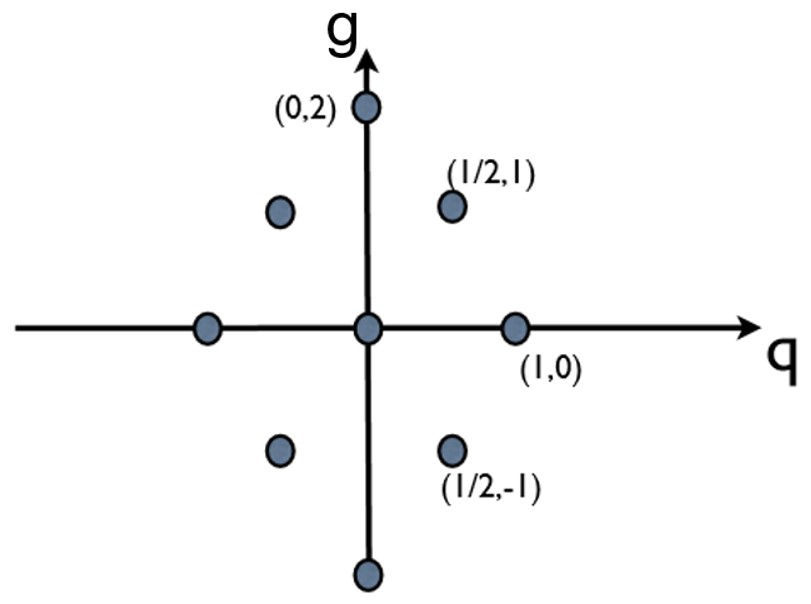}
 \caption{When $\theta$ is nonzero, the charge lattice is tilted.  The case $\theta=\pi$ is seen above.  (Adapted from Reference \onlinecite{chong}.)}
 \label{fig:thetapi}
 \end{figure}

We can very easily see this explicitly by introducing the auxiliary $\vec{E}$ field, as before, after which the action becomes:
\begin{equation}
S = \int \bigg((\vec{E}+\frac{\theta}{4\pi^2}\vec{B})\cdot(\dot{\vec{A}}-\vec{\nabla}A_0) - \frac{1}{2}E^2 - \frac{1}{2}B^2\bigg)
\end{equation}
After integrating the $A_0$ term by parts, then integrating $A_0$ out of the path integral, we see that the gauge constraint now becomes:
\begin{equation}
\vec{\nabla}\cdot\vec{E} = -\frac{\theta}{4\pi^2}\vec{\nabla}\cdot\vec{B}
\end{equation}
From the definition of the magnetic field, $\vec{B} = \vec{\nabla}\times\vec{A}$, and the compact nature of the gauge field, $A_i\sim A_i + 2\pi$, the divergence of the magnetic field is quantized as $\vec{\nabla}\cdot\vec{B} = 2\pi g$ for integer $g$.  The gauge constraint then indicates that any particle carrying magnetic charge $g$ must also carry an electric charge given by:
\begin{equation}
q = -\frac{\theta g}{2\pi} + n
\end{equation}
for integer $n$, which is the standard result of the Witten effect.\cite{witten}  (We must allow for the extra integer $n$, since a fundamental electric charge can always bind with the dyon.)  This attachment of electric charge to the monopole has the result of ``tilting" the charge lattice of the system, as seen in Figures \ref{fig:thetazero} and \ref{fig:thetapi}.

It is clear that the values of charge in the charge lattice return to their original values at $\theta = 2\pi$.  One might then naively conclude that $\theta$ is always a $2\pi$-periodic variable.  However, one more subtlety enters in the form of particle statistics.  For a system where the fundamental electric charge is a fermion, the naive expectation of $2\pi$-periodicity is indeed correct.  For a bosonic electric charge, on the other hand, the charge lattice has a different assignment of statistics at $\theta = 2\pi$ than at $\theta = 0$ (the ``statistical Witten effect").  In this case, the system turns out to be $4\pi$-periodic in $\theta$.\cite{spt,chong}

It is also important to note that, while $\theta$ can $a$ $priori$ take any value, the charge lattice can only be time-reversal invariant if $\theta = 0$ or $\theta = \pi$ (or multiples thereof), the two cases depicted in Figures \ref{fig:thetazero} and \ref{fig:thetapi}.  This is due to the fact that magnetic charge is odd under time-reversal, while electric charge is even, so a time-reversal invariant charge lattice must be symmetric about the $q$-axis.  It can further be shown that $\theta=\pi$ is only time-reversal invariant if the fundamental electric charge is a fermion, due to the effects of particle statistics.  These facts have important consequences both for symmetry protected topological phases protected by time-reversal symmetry \cite{spt} and for time-reversal invariant $U(1)$ spin liquids. \cite{chong}

\subsection{Vector Charge Theory}

For the vector charge theory, we derived that the action takes the form:
\begin{equation}
S = \int \bigg(\frac{1}{2}(\dot{A}_{ij} - \frac{1}{2}\partial_iC_j - \frac{1}{2}\partial_j C_i)^2 - \frac{1}{2}B^{ij}B_{ij}\bigg)
\label{vecact}
\end{equation}
with magnetic tensor $B_{ij} = \epsilon_{iab}\epsilon_{jcd}\partial^a\partial^c A^{bd}$.  We take $A_{ij}$ to be compact, $A_{ij}\sim A_{ij}+2\pi$, so the magnetic field admits monopoles of the form $\partial_iB^{ij} = 2\pi g^j$ for magnetic vector charge $g^j$, which is an integer linear combination of the fundamental charge vectors of the system.\cite{foot0}  As in Maxwell theory, we expect there to be some ``$E\cdot B$" term we can write down which attaches electric charge to these magnetic monopoles.  However, in this case, it will turn out that there are \emph{two} different types of $\theta$ terms that we can consider.

\subsubsection{The Direct Term}

The most natural analogue of an ``$E\cdot B$" term takes the form of a direct contraction between $E_{ij}$ and $B_{ij}$:
\begin{equation}
S_{\theta_1} = \frac{\theta_1}{4\pi^2} \int (\dot{A}_{ij} - \frac{1}{2}\partial_iC_j - \frac{1}{2}\partial_j C_i)B^{ij}
\end{equation}
\begin{equation}
= \frac{\theta_1}{4\pi^2} \int (\dot{A}_{ij} - \partial_iC_j)\epsilon^{iab}\epsilon^{jcd}\partial_a\partial_c A_{bd}
\end{equation}
Some algebra yields that this term is the integral of total derivatives, just like the conventional $\theta$ term:
\begin{align}
\begin{split}
S_{\theta_1} = \frac{\theta_1}{4\pi^2} \int \partial_a&(\dot{A}_{ij}\epsilon^{iab}\epsilon^{jcd}\partial_cA_{bd} + C_j \epsilon^{iab}\epsilon^{jcd}\partial_i\partial_c A_{bd})\\
&- \frac{1}{2}\partial_t(\epsilon^{iab}\partial_aA_{ij}\epsilon^{jcd}\partial_c A_{bd})
\end{split}
\end{align}
As such, this term has no effect on the bulk gauge mode.  However, it will contribute a boundary term which can result in the binding of charges to monopoles.  To see the charge attachment explicitly, we combine the $\theta_1$ term with the action of Equation \ref{vecact}.  After adding the auxiliary $E_{ij}$ field, the total action becomes:
\begin{align}
\begin{split}
S = \int (E^{ij}+\frac{\theta_1}{4\pi^2}B^{ij})(\dot{A}_{ij} - \partial_iC_j) \\
- \frac{1}{2}E^{ij}E_{ij} - \frac{1}{2}B^{ij}B_{ij}
\end{split}
\end{align}
After integrating by parts on the $C_j$ term, then integrating $C_j$ out of the path integral, we see that the gauge constraint becomes:
\begin{equation}
\partial_iE^{ij} = -\frac{\theta_1}{4\pi^2}\partial_iB^{ij}
\end{equation}
In terms of a particle's electric charge vector $\vec{q}$ and its magnetic charge vector $\vec{g}$, this equation tells us that:
\begin{equation}
\vec{q} = -\frac{\theta_1}{2\pi}\vec{g} + \vec{n}
\end{equation}
where $\vec{n}$ is any integer combination of the fundamental vector charges of the theory, representing the possibility of fundamental electric charges binding to the dyon.  We therefore see that the $\theta_1$ term has the effect of binding together \emph{parallel} electric and magnetic charge vectors.

It is easy to see that, when $\theta_1 = 2\pi$, the set of allowed charges will return to itself.  Furthermore, there are no concerns associated with statistics in this theory.  The particles in this case are 1-dimensional objects, for which there is no real distinction between bosons and fermions.  (One can check that there are also no statistical issues with bound states.)  We can therefore conclude that the $\theta_1$ parameter is indeed $2\pi$-periodic.  We can then further ask which values of $\theta_1$ are allowed if we enforce time-reversal invariance.  We note that, taking the convention that $\vec{q}$ is even under time reversal\cite{foot1}, $\vec{g}$ will be odd, transforming as $\vec{g}\rightarrow -\vec{g}$.  For the charge lattice to be invariant under this transformation, we then require:
\begin{equation}
-\frac{\theta_1}{2\pi}\vec{g} = \frac{\theta_1}{2\pi}\vec{g} + \vec{n}
\end{equation}
\begin{equation}
\Rightarrow -\frac{\theta_1}{\pi}\vec{g} = \vec{n}
\end{equation}
for some integer combination $\vec{n}$ of fundamental vector charges.  For this to be true, we require $\theta_1 = 0$ or $\theta_1 = \pi$, exactly as in the case of the conventional $\theta$ term.  Thus, just as in the usual $U(1)$ gauge theory, time-reversal invariance restricts the charge lattice of the theory to just two possible shapes.

\subsubsection{The Indirect Term}

\begin{figure}[t!]
 \centering
 \includegraphics[scale=0.35]{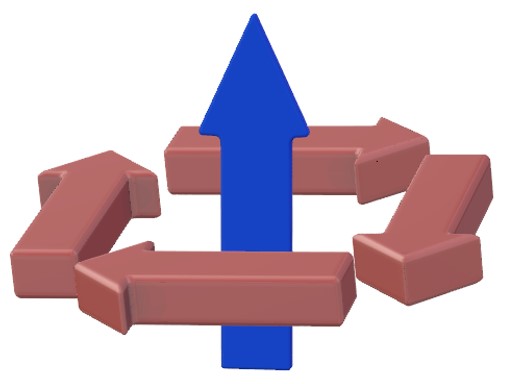}
 \caption{The $\theta_2$ term serves the purpose of attaching an angular configuration of electric charge (red) to the magnetic charge (blue).}
 \label{fig:angattach}
 \end{figure}

The $\theta_1$ term considered above led to a direct binding between the electric vector charges and their magnetic analogue.  However, the vector charge theory also hosts another class of nontrivial excitations: particles carrying zero net charge, but a nonzero angular charge moment.  The angular charge moment $\vec{\ell}$ of such a particle is also a vector quantity.  It therefore seems plausible that we could add a term which binds electric angular moment to the magnetic vector charge.  We can accomplish this by adding an ``indirect" contraction, of the form:
\begin{align}
\begin{split}
S_{\theta_2} = &\frac{\theta_2}{4\pi^2}\int \epsilon^{ik\ell}\partial_k(\dot{A}_{\ell j} - \frac{1}{2}\partial_\ell C_j - \frac{1}{2}\partial_jC_\ell ) B_i^{\,\,\,j} \\
&= \frac{\theta_2}{4\pi^2}\int \epsilon^{ik\ell}\partial_k(\dot{A}_{\ell j} - \frac{1}{2}\partial_jC_\ell ) B_i^{\,\,\,j}
\end{split}
\end{align}
Like the direct term, some tedious algebra reveals that the indirect $\theta_2$ term is also the integral of a total derivative:
\begin{align}
\begin{split}
S_{\theta_2} = \frac{\theta_2}{8\pi^2}\int &\partial_k(\epsilon^{ic\ell}\partial_c\dot{A}_{\ell j} \epsilon_{iab}\epsilon^{jkd}\partial^a A^b_{\,\,\,d} + \partial_iC_j\epsilon^{ji\ell}B_\ell^{\,\,\,k}) \\
&- \partial_t(\epsilon^{ik\ell}\partial_k\partial_c A_{\ell j} \epsilon_{iab}\epsilon^{jcd}\partial^aA^b_{\,\,\,d})
\end{split}
\end{align}
The gauge mode is therefore unaffected by the presence of a $\theta_2$ term.  Repeating the procedure from the previous section, introducing the auxiliary $E_{ij}$ and integrating out $C_i$, we find that the gauge constraint is:
\begin{equation}
\partial_i E^{ij} = -\frac{\theta_2}{8\pi^2}\epsilon^{jk\ell}\partial_k \partial_iB_\ell^{\,\,\,i}
\end{equation}
In terms of electric and magnetic charge vectors $\vec{q}$ and $\vec{g}$, this implies:
\begin{equation}
\vec{q} = -\frac{\theta_2}{4\pi}(\vec{\nabla}\times\vec{g}) + \vec{n}
\end{equation}
where $\vec{n}$ is any integer combination of the fundamental charge vectors.  The physical consequence of this is that, for every magnetic vector charge, there is a configuration of electric charges winding around perpendicular to it, as seen in Figure \ref{fig:angattach}.  In other words, the magnetic monopoles have electric angular charge attached, as promised.

The set of allowed charges will return to itself only at $\theta_2 = 4\pi$ (unlike the $2\pi$ repetition of $\theta_1$).  Furthermore, the attachment of electric angular charge leaves the magnetic vectors as one-dimensional particles, for which statistics are not meaningful.  We therefore conclude that $\theta_2$ is a $4\pi$-periodic variable.  Similarly, it is straightforward to check that $\theta_2 = 0$ and $\theta_2 = 2\pi$ are the only values allowed in the presence of time-reversal invariance.

\subsection{Scalar Charge Theory}

We now move on to apply the same principles to the scalar charge theory.  Since there is no self-duality in this case, there is no term which can attach an electric particle to its magnetic analogue.  The electric charges of the theory are scalars, while the magnetic charges are vectors.  The most natural form of charge attachment in this theory is attaching an electric dipole moment to a magnetic vector.  Indeed, we will find that the direct contraction between $E^{ij}$ and $B^{ij}$ accomplishes precisely this.

For the scalar charge theory, the action takes the form:
\begin{equation}
S = \int\bigg(\frac{1}{2}(\dot{A}_{ij} - \partial_i\partial_j\phi)^2 - \frac{1}{2}B^{ij}B_{ij}\bigg)
\end{equation}
where the magnetic tensor is $B_{ij} = \epsilon_{iab}\partial^a A^b_{\,\,\,j}$, admitting monopoles of the form $\partial_i B^{ij} = 2\pi g^j$ for magnetic vector charge $g^j$ (an integer linear combination of fundamental charge vectors).  Note the distinctly different behavior of the electric and magnetic sectors.  The electric tensor $E^{ij}$ is symmetric and is characterized by scalar fracton charges.  The magnetic tensor $B^{ij}$, on the other hand, is traceless and non-symmetric, characterized by vector charges, which can be shown to behave as 2-dimensional particles, moving only transversely to their charge vector.\cite{sub}

\begin{figure}[b!]
 \centering
 \includegraphics[scale=0.35]{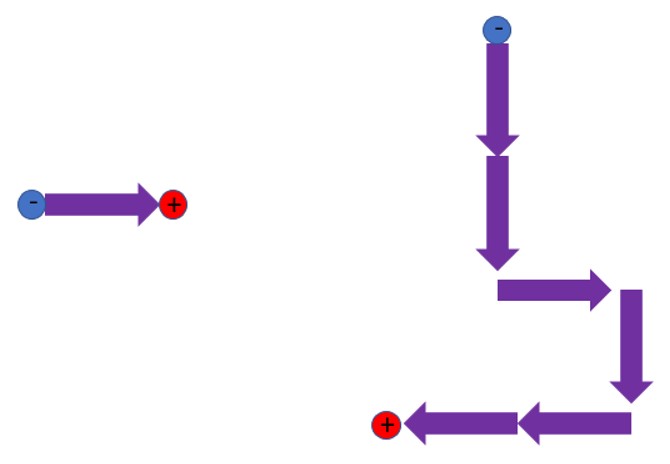}
 \caption{When $\theta$ is nonzero, electric fractons will be attached to the endpoints of the magnetic charge vectors.  Due to the finite gap to electric charges, it will be energetically favorable for the magnetic charges to bind into string-like objects.}
 \label{fig:attach}
 \end{figure}

Despite the lack of self-duality in the theory, we can still write down a natural ``$E\cdot B$" term and investigate its consequences:
\begin{equation}
S_\theta = \frac{\theta}{4\pi^2}\int (\dot{A}_{ij} - \partial_i\partial_j\phi)\epsilon^{iab}\partial_aA_{b}^{\,\,\,j}
\end{equation}
Simple algebra yields that this only contributes a boundary term to the action:
\begin{align}
\begin{split}
S_\theta = \frac{\theta}{4\pi^2}\int \partial_a(\frac{1}{2}\epsilon^{iab}\dot{A}_{ij} A_b^{\,\,\,j}+\partial_j\phi\epsilon^{iab}\partial_i A_b^{\,\,\,j})\\
- \frac{1}{2}\partial_t(\epsilon^{iab}\partial_aA_{ij}A_b^{\,\,\,j})
\end{split}
\end{align}
so once again this $\theta$ term will have no consequences for the bulk gauge mode of the system.  It will, however, introduce boundary physics which will have consequences for the charges of the system.  Once again, we can see this by introducing the auxiliary $E_{ij}$ field, upon which the total action becomes:
\begin{align}
\begin{split}
S = \int (E^{ij}+\frac{\theta}{4\pi^2}B^{ij})(\dot{A}_{ij}-\partial_i\partial_j\phi)\\
-\frac{1}{2}E^{ij}E_{ij} - \frac{1}{2}B^{ij}B_{ij}
\end{split}
\end{align}
After integrating by parts on the $\phi$ term and integrating $\phi$ out of the path integral, our gauge constraint becomes:
\begin{equation}
\partial_i\partial_j E^{ij} = -\frac{\theta}{4\pi^2}\partial_i\partial_jB^{ij}
\end{equation}
In terms of the electric scalar charge $\partial_i\partial_j E^{ij} = q$ and the magnetic vector charge $\partial_iB^{ij} = 2\pi g^j$, this equation becomes:
\begin{equation}
q = -\frac{\theta}{2\pi}\partial_ig^i + n
\end{equation}
for integer $n$.  In this case, the $\theta$ term has the effect of attaching fractonic electric charges to the \emph{endpoints} of the magnetic charge vectors, as seen in Figure \ref{fig:attach}, with opposite sign charges at each end of the vector.  Equivalently, we can think of the $\theta$ term as binding an electric dipole to each magnetic vector.  We can then see that, if we construct a long directed string of such vector charges, we will only have electric charges at the two endpoints of the string.  Since it takes finite energy to create the electric charges, it will be energetically favorable for a group of magnetic vectors to line up into such string arrangements, so as to minimize the number of electric charges present.

We therefore see that, when $\theta\neq 0$, we can more usefully think of the system in terms of a theory of open strings, as opposed to point particles (though there is strictly speaking no rigid distinction between the two pictures).  And in the limit where the gap to electric charges goes to infinity, we can understand the behavior of the magnetic charges in terms of a theory of closed strings (coupled to a tensor gauge field).  If all values of $\theta$ are regarded as equally likely, then $\theta = 0$ is a somewhat special case where the magnetic charges have no tendency to line up.  For any nonzero $\theta$, the magnetic charges will arrange themselves into open string configurations to minimize electric charge.

As in the previous theories, we see that $\theta = 2\pi$ corresponds to full integer multiples of charge being attached to the endpoints of magnetic vectors, which takes us back to the original set of allowed charges.  However, in this case, we must also worry about particle statistics.  Note that a magnetic vector charge has mutual $\pi$ statistics with a parallel electric dipole in its plane of motion (see the Appendix).  Assuming the magnetic vectors and electric dipoles are both bosons\cite{foot2} (as is the case in the simplest lattice models\cite{alex,sub}), then the bound state of the two will be a fermion.  It is then easy to check that, when $\theta = 2\pi$, the charge lattice has returned to its original shape, but the assignment of statistics is different.  Only at $\theta = 4\pi$ does the system go back to the original charge lattice with the original statistics.  We therefore see that $\theta$ is $4\pi$-periodic in this system.  By similar logic, it is straightforward to check that imposing time-reversal invariance will restrict the $\theta$ parameter to only two possible distinct values: $\theta = 0$ or $\theta = 2\pi$.

\section{Two-Dimensional Fracton Phases}

In the above analysis, we found that $\theta$ terms in the action for the higher-spin $U(1)$ gauge theories only lead to boundary contributions.  This is the same situation as in Maxwell theory, where a $\theta$ term in the bulk leads to a Chern-Simons term on the boundary of the system.  The study of such Chern-Simons terms is important not only for understanding the boundary of three-dimensional $U(1)$ spin liquids, but also for shedding light on quantum hall physics in purely two-dimensional systems.  We therefore wish to study the boundary terms which arise as a consequence of $\theta$ terms in the higher-spin $U(1)$ gauge theories.  These terms are obviously important for studying the boundary of tensor $U(1)$ spin liquids.  But perhaps even more importantly, they will teach us how to formulate fundamentally new two-dimensional phases of matter supporting fracton excitations.  We will comment later on how these phases get around the conventional folklore that fractons do not occur in two-dimensional systems.

\subsection{Review of Chern-Simons Theory}

We first review a perspective on the Chern-Simons boundary terms in Maxwell theory with a $\theta$ term, so as to facilitate comparison with the higher-spin case.  For Maxwell theory, the $\theta$ term can be written as:
\begin{align}
\begin{split}
S_\theta = \frac{\theta}{4\pi^2}\int (\dot{A}_i& - \partial_i A_0)\epsilon^{ijk}\partial_jA_k\\
=-\frac{\theta}{4\pi^2}\int& \partial_i(A_0\epsilon^{ijk}\partial_jA_k + \frac{1}{2}\dot{A}_j\epsilon^{ijk}A_k)\\
&+ \frac{1}{2}\partial_t(\partial_jA_i\epsilon^{ijk}A_k)
\end{split}
\end{align}
For a spatial boundary, taken with normal $\hat{z}$ for simplicity, the boundary action is then:
\begin{equation}
S_\partial = -\frac{\theta}{4\pi^2}\int_\partial A_0\epsilon^{jk}\partial_jA_k + \frac{1}{2}\epsilon^{jk}\dot{A}_jA_k
\end{equation}
where the integral is over the boundary.  This is the usual Chern-Simons theory, expanded out in terms of spatial and timelike components.  The first term gives us the following constraint on the low-energy sector:
\begin{equation}
-\frac{\theta}{4\pi^2}\epsilon^{jk}\partial_jA_k = -\frac{\theta}{4\pi^2}B = 0
\end{equation}
where $B$ is the magnetic flux through the surface.  More generally, allowing for charges coupled to the Chern-Simons field, we will have:
\begin{equation}
\rho = -\frac{\theta}{4\pi^2}B
\end{equation}
for charge density $\rho$, which indicates that each $2\pi$ flux has $-\theta/2\pi$ electric charge attached.  (Or equivalently, each charge has $-4\pi^2/\theta$ flux attached.)  Note that $\rho$ is a divergence, $\rho = \partial_j(-\theta\epsilon^{jk}A_k/4\pi^2)$, so the usual charge conservation law will hold.  Conservation of this charge on the two-dimensional boundary is a direct consequence of the fact that magnetic flux through the boundary cannot change unless a magnetic monopole is passed through the surface into the three-dimensional bulk.

The second term in the boundary action, $\frac{1}{2}\epsilon^{jk}\dot{A}_jA_k$, determines the canonical structure of the gauge field, dictating that $A_x$ and $A_y$ are canonical conjugates, as is familiar in Chern-Simons theory.  The presence of only one time derivative leads to the vanishing of these terms upon going to the Hamiltonian formalism, indicating that the Hamiltonian itself vanishes within the low-energy gauge sector, reflecting the lack of local degrees of freedom of a topological field theory.

To verify that the Chern-Simons theory has a finite energy gap, it is instructive to regularize the theory by including a two-dimensional Maxwell term, after which the action becomes:
\begin{align}
\begin{split}
S = \int \bigg(\frac{1}{2e^2}((\dot{A}_i - \partial_iA_0)^2 - (\epsilon^{ij}\partial_iA_j)^2) \\
-\frac{\theta}{4\pi^2}(A_0\epsilon^{ij}\partial_iA_j + \frac{1}{2}\epsilon^{ij}\dot{A}_iA_j)\bigg)
\end{split}
\end{align}
Varying the action with respect to $A_i$, the resulting equation of motion is:
\begin{equation}
\frac{1}{e^2}\dot{E}^i + \frac{\theta}{4\pi^2}\epsilon^{ij}E_j - \frac{1}{e^2}\epsilon^{ij}\partial_jB = 0
\end{equation}
At low energies, the last term is irrelevant, and we obtain an energy gap scaling as $\Delta \sim \theta e^2$.  As we take $e\rightarrow\infty$, the action approaches pure Chern-Simons theory, and the energy gap approaches infinity.

When the Chern-Simons term is regarded as an effective action for the physical external electromagnetic gauge field of a two-dimensional system, we can easily derive the Hall response:
\begin{equation}
\langle J^i\rangle = -\frac{\delta S}{\delta A_i} = -\frac{\theta}{4\pi^2}\epsilon^{ij}\dot{A}_j = -\frac{\theta}{4\pi^2}\epsilon^{ij}E_j
\end{equation}
so the conductivity tensor is:
\begin{equation}
\sigma^{ij} = -\frac{\theta}{4\pi^2}\epsilon^{ij}
\end{equation}
If we restrict $\theta$ to be a multiple of $2\pi$, such that there are no fractionalized charges in the system, we see that the Hall response must be an integer multiple of some fundamental value ($1/2\pi$ in the conventions taken here), representing the integer quantum Hall effect.  These correspond to the ``properly" quantized levels of the Chern-Simons theory.  For other values of $\theta$, with improperly quantized Chern-Simons terms, we will have both a fractional Hall response and fractional charges on the boundary, in a manifestation of the fractional quantum Hall effect.

\subsection{Boundary of the Vector Charge Theory}

Let us now carry these same principles over to a higher-spin $U(1)$ gauge theory, namely the vector charge theory.  We will apply the analysis separately to the direct $\theta_1$ term and the indirect $\theta_2$ term.

\subsubsection{The Direct Term}

We found that the $\theta_1$ term can be written in the following total derivative form:
\begin{align}
\begin{split}
S_{\theta_1} = \frac{\theta_1}{4\pi^2}\int \partial_a&(\dot{A}_{ij}\epsilon^{iab}\epsilon^{jcd}\partial_cA_{bd} + C_j\epsilon^{iab}\epsilon^{jcd}\partial_i\partial_cA_{bd})\\
&- \frac{1}{2}\partial_t(\epsilon^{iab}\partial_aA_{ij}\epsilon^{jcd}\partial_cA_{bd})
\end{split}
\end{align}
Taking a spatial boundary with normal $\hat{z}$, the boundary action becomes:
\begin{equation}
S_\partial = \frac{\theta_1}{4\pi^2}\int_\partial \dot{A}_{ij}\epsilon^{bi}\epsilon^{jcd}\partial_cA_{bd} + C_j\epsilon^{bi}\epsilon^{jcd}\partial_i\partial_cA_{bd}
\end{equation}
where we have introduced the 2d Levi-Civita symbol $\epsilon^{ij}$ in the $xy$-plane.  From the second term, we can see that the Lagrange multiplier $C_i$ imposes several constraints on the boundary theory:
\begin{equation}
\frac{\theta_1}{4\pi^2}\epsilon^{bi}\epsilon^{cd}\partial_i\partial_c A_{bd} = 0
\end{equation}
\begin{equation}
\frac{\theta_1}{4\pi^2}\epsilon^{bi}\epsilon^{jc}\partial_i(\partial_c A_{bz} -\partial_z A_{bc})=0
\end{equation}
Right away, we can see that this theory is going to be a bit more complicated than the conventional Chern-Simons theory.  In that case, the $A_z$ component and all $z$ derivatives were eliminated in the boundary theory.  In the present case, on the other hand, only $A_{zz}$ is completely eliminated from the theory, while $A_{iz}$ (for $i\neq z$) and $z$ derivatives remain.  In order to handle this, we first break up the three-dimensional tensor $A_{ij}$ as follows:
\begin{equation}
a_{ij} = A_{ij}\,\,\,\textrm{for}\,\,i,j=x,y\,\,\,\,\,(\textrm{a 2d tensor)}
\end{equation}
\begin{equation}
\lambda_i = A_{iz}\,\,\,\textrm{for}\,\,i = x,y\,\,\,\,\,(\textrm{a 2d vector})
\end{equation}
We also note that $a_{ij}$ and $\partial_z a_{ij}$ can be independently varied on the boundary of the system.  We therefore define another independent symmetric tensor field as:
\begin{equation}
\gamma_{ij} = -\partial_za_{ij}\,\,\,\,\,(\textrm{a 2d tensor})
\end{equation}
In terms of the three fields $a_{ij}$, $\lambda_i$, and $\gamma_{ij}$, the constraint equations become:
\begin{equation}
\frac{\theta_1}{4\pi^2}\epsilon^{bi}\epsilon^{cd}\partial_i\partial_c a_{bd} = 0
\end{equation}
\begin{equation}
\frac{\theta_1}{4\pi^2}\epsilon^{bi}\epsilon^{jc}\partial_i(\partial_c \lambda_b + \partial_b\lambda_c + \gamma_{bc}) = 0
\label{symm}
\end{equation}
(The middle term in Equation \ref{symm} is identically zero, but has been added for later convenience.)  In terms of these fields, the dynamic part of the action (involving time derivatives) becomes:
\begin{align}
\begin{split}
S_{dyn} = \frac{\theta_1}{4\pi^2}&\int_\partial \dot{a}_{ij}\epsilon^{bi}\epsilon^{jc}\partial_c\lambda_b + \dot{a}_{ij}\epsilon^{bi}\epsilon^{jd}\gamma_{bd} + \dot{\lambda}_i\epsilon^{bi}\epsilon^{cd}\partial_ca_{bd}\\
&= \frac{\theta_1}{4\pi^2}\int_\partial \dot{a}_{ij}\epsilon^{bi}\epsilon^{jc}(\partial_c\lambda_b + \partial_b\lambda_c + \gamma_{bc})
\end{split}
\end{align}
From the constraints and dynamic terms above, it becomes clear that we can eliminate $\lambda_i$ entirely and simplify the theory by shifting:
\begin{equation}
\gamma_{ij}\rightarrow\gamma_{ij} - \partial_i\lambda_j - \partial_j\lambda_i
\end{equation}
In terms of the remaining two fields $a_{ij}$ and $\gamma_{ij}$, the dynamical action takes the form:
\begin{equation}
S_{dyn}[a_{ij},\gamma_{ij}] = \frac{\theta_1}{4\pi^2}\int_\partial \dot{a}_{ij}\epsilon^{bi}\epsilon^{jc}\gamma_{bc}
\end{equation}
which is subject to the following constraints:
\begin{equation}
\frac{\theta_1}{4\pi^2}\epsilon^{bi}\epsilon^{jc}\partial_i\partial_j a_{bc} = 0
\end{equation}
\begin{equation}
\frac{\theta_1}{4\pi^2}\epsilon^{bi}\epsilon^{jc}\partial_i\gamma_{bc} = 0
\end{equation}
Note that the six independent components of the pair $a_{ij},\gamma_{ij}$ are organized into three canonical conjugate pairs.  Since we have three independent constraints on the system, we see that the pure gauge theory is fully constrained, with no local degrees of freedom.  It should be noted that this theory is a tensor generalization of a mutual Chern-Simons theory and is non-chiral.  (The action is invariant under spatial reflections.)

More generally, we can consider charged defects coupled to the gauge field as sources for the constraint equations:
\begin{equation}
\frac{\theta_1}{4\pi^2}\epsilon^{bi}\epsilon^{jc}\partial_i\partial_j a_{bc} = \rho
\end{equation}
\begin{equation}
\frac{\theta_1}{4\pi^2}\epsilon^{bi}\epsilon^{jc}\partial_i \gamma_{bc} = \pi^j
\end{equation}
for scalar charge density $\rho$ and vector charge density $\pi^j$.  From the general arguments regarding Gauss's laws and conservation laws\cite{sub}, we can see that the scalar charge has a conserved dipole moment:
\begin{equation}
\int d^2x\,(\rho\vec{x}) = \textrm{constant}
\end{equation}
This restricts these charges from moving in any direction, meaning that the charge $\rho$ represents fracton excitations.  Similarly, it is not hard to show that $\pi^j$ represents one-dimensional particles, only moving in the direction of their charge vector.  Intuitively, we can understand these results from the fact that electric vector charges are bound to magnetic flux on the boundary.  Electric charge vectors in the $xy$-plane provide 1-dimensional particles on the boundary, while charge vectors in the $z$ direction have no free motion at all on the boundary and give us fractons in the boundary theory.

It should be noted that being in a system with $U(1)$ charge behavior, where charge is conserved absolutely, is crucial to the fracton structure.  In a discrete charge system, where charge is only conserved modulo some integer, it is known that these sorts of conservation laws break down in two dimensions\cite{fracton1,fracton2}, and the particles become fully mobile.

To verify that there is a finite energy gap, we once again regularize the theory by keeping finite two-dimensional Maxwell terms for the two gauge fields, after which the action becomes:
\begin{align}
\begin{split}
S = \int\bigg(\frac{1}{2e^2}((\dot{a}_{ij} &- \frac{1}{2}(\partial_iC_j+\partial_jC_i))^2 - (\epsilon^{ij}\epsilon^{k\ell}\partial_i\partial_ka_{j\ell})^2) \\
+ \frac{1}{2e^2}((&\dot{\gamma}_{ij} - \partial_i\partial_j\phi)^2 - (\epsilon^{jk}\partial_j\gamma_{ki})^2)\\
 + \frac{\theta_1}{4\pi^2}(\dot{a}_{ij}\epsilon^{bi}\epsilon^{jc}&\gamma_{bc} +\phi\epsilon^{bi}\epsilon^{jc}\partial_i\partial_ja_{bc} + C_j\epsilon^{jc}\epsilon^{bi}\partial_i\gamma_{bc}     )\bigg)
\end{split}
\end{align}
(We have written a vector charge Maxwell term for $a_{ij}$ and a scalar charge Maxwell term for $\gamma_{ij}$, which is necessary in order to match the charge structure of the Chern-Simons theory.)  Varying with respect to the gauge fields, the equations of motion are:
\begin{equation}
\frac{1}{e^2}\dot{E}_{(a)}^{ij} - \frac{\theta_1}{4\pi^2}\epsilon^{bi}\epsilon^{jc}E^{(\gamma)}_{bc} - \frac{1}{e^2}\epsilon^{ik}\epsilon^{j\ell}\partial_k\partial_\ell B_{(a)} = 0
\end{equation}
\begin{equation}
\frac{1}{e^2}\dot{E}^{ij}_{(\gamma)} + \frac{\theta_1}{4\pi^2}\epsilon^{bi}\epsilon^{jc}E_{bc}^{(a)} - \frac{1}{e^2}\epsilon^{ik}\partial_kB^j_{(\gamma)} = 0
\end{equation}
As in conventional Chern-Simons theory, the magnetic terms in the above equations are irrelevant at low energies, and all gauge modes have a gap scaling as $\Delta\sim \theta_1 e^2$.  As $e\rightarrow\infty$, we recover a pure tensor Chern-Simons theory, and the gap approaches infinity.  Therefore, the tensor Chern-Simons theory describes a stable gapped phase of matter.

We can use our new Chern-Simons action to derive a generalized ``Hall" response of the system.  If we have a current $J^{ij}_{(a)}$ coupled to the action via $J^{ij}_{(a)}a_{ij}$, and similarly for $\gamma_{ij}$, then the generalized Hall responses are:
\begin{equation}
\langle J^{ij}_{(a)}\rangle = - \frac{\partial S}{\partial a_{ij}} = \frac{\theta_1}{4\pi^2}\epsilon^{bi}\epsilon^{jc}\dot{\gamma}_{bc} = \frac{\theta_1}{4\pi^2}\epsilon^{bi}\epsilon^{jc}E^{(\gamma)}_{bc}
\end{equation}
\begin{equation}
\langle J^{ij}_{(\gamma)}\rangle = -\frac{\partial S}{\partial \gamma_{ij}} = \frac{\theta_1}{4\pi^2}\epsilon^{bi}\epsilon^{jc}\dot{a}_{bc} = \frac{\theta_1}{4\pi^2}\epsilon^{bi}\epsilon^{jc}E^{(a)}_{bc}
\end{equation}
Note that these do not represent responses to an externally applied electric field, but rather to the internal emergent tensor field.  This is more difficult to control from an experimental standpoint.  Nevertheless, large-scale configurations of the emergent tensor field can in principle be set up, perhaps through applying appropriate strains on the system.  In such a scenario, the generalized Hall response should be a measurable quantity.  Furthermore, when there are no fractional charges in the system, $\theta_1$ will be restricted to a multiple of $2\pi$, and the Hall responses will be integer multiples of a fundamental unit, in close analogy with the integer quantum Hall effect.  Values of $\theta_1$ other than these ``properly" quantized levels will lead to fractional generalized Hall responses and also fractional charges.

\subsubsection{The Indirect Term}

We found that the $\theta_2$ term can be written as:
\begin{align}
\begin{split}
S_{\theta_2} = \frac{\theta_2}{8\pi^2}\int &\partial_k(\epsilon^{ic\ell}\partial_c\dot{A}_{\ell j} \epsilon_{iab}\epsilon^{jkd}\partial^a A^b_{\,\,\,d} + \partial_iC_j\epsilon^{ji\ell}B_\ell^{\,\,\,k}) \\
&- \partial_t(\epsilon^{ik\ell}\partial_k\partial_c A_{\ell j} \epsilon_{iab}\epsilon^{jcd}\partial^aA^b_{\,\,\,d})
\end{split}
\end{align}
Taking a spatial boundary with normal $\hat{z}$, the boundary action becomes:
\begin{equation}
S_{\theta_2} = \frac{\theta_2}{8\pi^2}\int_\partial (\epsilon^{ic\ell}\partial_c\dot{A}_{\ell j}\epsilon_{iab}\epsilon^{jzd}\partial^a A^b_{\,\,\,d} + \partial_iC_j\epsilon^{ji\ell}B_\ell^{\,\,\,z})
\end{equation}
We note that, since $\partial_zC_j$ can be varied independently of $C_j$ on the boundary, the theory now has two extra Lagrange multipliers (there is no constraint from $\partial_zC_z$).  Integrating out all such Lagrange multipliers, our system obeys five total constraints:
\begin{equation}
\frac{\theta_2}{8\pi^2}\epsilon^{cd}\partial_c(\partial^j A^z_{\,\,\,d} - \partial^z A^j_{\,\,\,d}) = 0
\end{equation}
\begin{equation}
\frac{\theta_2}{8\pi^2}\epsilon^{cd}\partial_i\partial_c(\partial^z A^i_{\,\,\,d} - \partial^i A^z_{\,\,\,d}) = 0
\end{equation}
\begin{equation}
\frac{\theta_2}{8\pi^2}\epsilon^{ji}\epsilon^{ab}\epsilon^{cd}\partial_i\partial_a\partial_c A_{bd} = 0
\end{equation}
The second constraint is simply the divergence of the first and is therefore redundant.  In terms of the decomposition of the previous section, the independent constraints are:
\begin{equation}
\frac{\theta_2}{8\pi^2}\epsilon^{cd}\partial_c(\gamma^{jd}+\partial^j \lambda^d) = 0
\end{equation}
\begin{equation}
\frac{\theta_2}{8\pi^2}\epsilon^{ji}\epsilon^{ab}\epsilon^{cd}\partial_i\partial_a\partial_c a_{bd} = 0
\end{equation}
and the dynamical piece of the action can be written as:
\begin{align}
\begin{split}
S_{dyn} = \frac{\theta_2}{8\pi^2}\int_\partial \epsilon^{ij}(\epsilon^{ab}\partial_a \dot{a}_{bi}\epsilon^{cd}\partial_ca_{dj}) \\
+ \epsilon^{ij}(\dot{\gamma}_{ai} + \partial_a\dot{\lambda}_i)(\gamma^a_{\,\,\,j} + \partial^a\lambda_j)
\end{split}
\end{align}
We shift $\gamma^{ij}\rightarrow\gamma^{ij} - \partial^i\lambda^j$, after which $\gamma^{ij}$ becomes a \emph{non-symmetric} tensor.  After the change of variables, we can write the constraints in the form:
\begin{equation}
\frac{\theta_2}{8\pi^2}\epsilon_{ab}\partial^a\gamma^{jb} = 0
\end{equation}
\begin{equation}
\frac{\theta_2}{8\pi^2}\epsilon^{ji}\epsilon^{ab}\epsilon^{cd}\partial_i\partial_a\partial_c a_{bd} = 0
\end{equation}
and the dynamical action becomes:
\begin{equation}
S_{dyn} = \frac{\theta_2}{8\pi^2}\int_\partial \epsilon^{ij}(\epsilon^{ab}\partial_a \dot{a}_{bi}\epsilon^{cd}\partial_ca_{dj}) + \epsilon^{ij}\dot{\gamma}_{ai}\gamma^a_{\,\,\,j}
\end{equation}
We can now see that the boundary decomposes into two independent theories.  One is a theory of a non-symmetric tensor $\gamma^{ij}$, described by the following action and constraint:
\begin{equation}
S_1 = \frac{\theta_2}{8\pi^2}\int_\partial\epsilon^{ij}\dot{\gamma}_{ai}\gamma^a_{\,\,\,j}
\end{equation}
\begin{equation}
\frac{\theta_2}{8\pi^2}\epsilon_{ab}\partial^a\gamma^{jb} = 0
\end{equation}
The four components of $\gamma^{ij}$ come together to form two independent degrees of freedom, described by a chiral generalized Chern-Simons theory.  Since we have two constraints on the system, we see that the system is fully constrained.  If we introduce charges, via $\rho^j = (\theta_2/8\pi^2)\epsilon_{cd}\partial^c \gamma^{jd}$, we will find vector charges that are fully mobile.  (The tensor $\gamma^{ij}$ is neither traceless nor symmetric, so there are no extra conservation laws beyond conservation of total charge.)  We can easily find the generalized Hall response to be:
\begin{equation}
\langle J^{ij}_{(\gamma)}\rangle = -\frac{\partial S}{\partial\gamma_{ij}} = \frac{\theta_2}{8\pi^2}\epsilon^{j}_{\,\,\,c}\dot{\gamma}^{ci} = \frac{\theta_2}{8\pi^2}\epsilon^{j}_{\,\,\,c}E^{ci}_{(\gamma)}
\end{equation}
To demonstrate the energy gap, we regularize with a Maxwell term:
\begin{align}
\begin{split}
S = \int \bigg(\frac{1}{2e^2}\bigg[(\dot{\gamma}^{ij}& - \frac{1}{2}(\partial_iC_j + \partial_jC_i))^2 - (\epsilon^{ik}\epsilon^{j\ell}\partial_i\partial_j \gamma_{k\ell})^2\bigg] \\
&+\frac{\theta_2}{8\pi^2}(\epsilon^{ij}\dot{\gamma}_{ik}\gamma_j^{\,\,\,k} + C_j\epsilon_{ik}\partial^i\gamma^{jk})\bigg)
\end{split}
\end{align}
As usual, the equation of motion behaves as $\frac{1}{e^2}\dot{E}_{(\gamma)}\sim \theta_2 E_{(\gamma)}$ at low energies, giving a gap which behaves as $\Delta\sim \theta_2 e^2$.

Meanwhile, the other decoupled theory on the boundary is a theory of a symmetric tensor $a^{ij}$, described by:
\begin{equation}
S_2 = \frac{\theta_2}{8\pi^2}\int_\partial \epsilon^{ij}(\epsilon^{ab}\partial_a \dot{a}_{bi}\epsilon^{cd}\partial_ca_{dj})
\end{equation}
\begin{equation}
\frac{\theta_2}{8\pi^2}\epsilon^{ji}\epsilon^{ab}\epsilon^{cd}\partial_i\partial_a\partial_c a_{bd} = 0
\label{constr}
\end{equation}
In this case, two of the components of $a^{ij}$ come together to form one canonical conjugate pair, giving a single degree of freedom.  (There is a third independent component of $a_{ij}$ which makes no appearance in either the action or constraint, and can therefore be eliminated.)  The constraint of Equation \ref{constr} is enough to fully constrain this single degree of freedom, eliminating any local degrees of freedom.  We can also readily check that this theory is chiral.  We can introduce a vector-valued charge via $\rho^j = (\theta_2/8\pi^2)\epsilon^{ji}\epsilon^{ab}\epsilon^{cd}\partial_i\partial_a\partial_c a_{bd}$.  Due to the high number of derivatives, these vector charges will be fracton excitations.  The generalized Hall response of the system is given by:
\begin{equation}
\langle J^{ij}_{(a)}\rangle = \frac{\theta_2}{8\pi^2}\epsilon^{ic}\epsilon^{jd}\epsilon^{ab}\partial_a\partial_c\dot{a}_{bd} = \frac{\theta_2}{8\pi^2}\epsilon^{ic}\epsilon^{jd}\epsilon^{ab}\partial_a\partial_cE_{bd}
\end{equation}

It is important to note that, while all previous Chern-Simons theories were gapped and manifestly stable, the theory describing $a_{ij}$ is not so lucky.  Let us regularize the theory with a Maxwell term, as usual:
\begin{align}
\begin{split}
S = \int \bigg(\frac{1}{2e^2}\bigg[(\dot{a}^{ij} - \frac{1}{2}(\partial_iC_j + \partial_jC_i))^2 - (\epsilon^{ik}\epsilon^{j\ell}\partial_i\partial_ja_{k\ell})^2\bigg] \\
+ \frac{\theta_2}{8\pi^2}(\epsilon^{ij}\epsilon^{ab}\partial_a\dot{a}_{bi}\epsilon^{cd}\partial_ca_{dj} + C_j\epsilon^{ji}\epsilon^{ab}\epsilon^{cd}\partial_i\partial_a\partial_c a_{bd})\bigg)
\end{split}
\end{align}
The equation of motion takes the form:
\begin{equation}
\frac{1}{e^2}\dot{E}^{ij} - \frac{\theta_2}{8\pi^2}\epsilon^{ic}\epsilon^{kj}\epsilon^{ab}\partial_c\partial_a E_{bk} + \frac{1}{e^2}\epsilon^{ik}\epsilon^{j\ell}\partial_k\partial_\ell B = 0
\end{equation}
Contracting the above equation with $\epsilon_{in}\epsilon_{jm}\partial^n\partial^m$, and taking advantage of Gauss's law, we can also write an equation of motion for $B$ as:
\begin{equation}
\frac{1}{e^2}\ddot{B} + \frac{1}{e^2}\partial^4 B = 0
\end{equation}
which is unchanged from the bare Maxwell theory, giving an $\omega\sim k^2$ dispersion.  Unlike the other tensor Chern-Simons terms we have considered, this term does not produce an energy gap, and therefore cannot stabilize the two-dimensional gauge field.  This is related to the fact that the Maxwell and Chern-Simons terms have the same number of derivatives, so there is no reason for pure Chern-Simons to dominate at low energies.  The Chern-Simons theory for $a_{ij}$ will have the same confinement instability as the pure Maxwell theory\cite{alex}, so this gauge field does not describe a stable two-dimensional phase of matter.  This instability will not affect the other gauge field, $\gamma_{ij}$, which decouples from $a_{ij}$ and remains in a stable deconfined phase.

\subsection{Boundary of the Scalar Charge Theory}

We next apply similar logic to the scalar charge theory, where we found that the $\theta$ term takes the form:
\begin{align}
\begin{split}
S_{\theta} = \frac{\theta}{4\pi^2}\int \partial_a(\frac{1}{2}\epsilon^{iab}\dot{A}_{ij}A_b^{\,\,\,j} + \partial_j\phi\epsilon^{iab}\partial_iA_b^{\,\,\,j}) \\
-\frac{1}{2}\partial_t(\epsilon^{iab}\partial_aA_{ij}A_b^{\,\,\,j})
\end{split}
\end{align}
For a spatial boundary with normal in the $\hat{z}$ direction, the action becomes:
\begin{equation}
S_\partial = \frac{\theta}{4\pi^2}\int_\partial \frac{1}{2}\epsilon^{bi}\dot{A}_{ij}A_b^{\,\,\,j} + \partial_j\phi\epsilon^{bi}\partial_iA_b^{\,\,\,j}
\end{equation}
We consider the second term first.  The $z$ derivative of $\phi$ can be varied independently of $\phi$ itself on the boundary and constitutes an independent Lagrange multiplier, while the in-plane components of $\partial_j\phi$ can be integrated by parts to yield a second constraint equation.  Using the decomposition from the previous sections, the two constraints take the form:
\begin{equation}
-\frac{\theta}{4\pi^2}\epsilon^{bi}\partial_i\partial_ja_b^{\,\,\,j} = 0
\end{equation}
\begin{equation}
\frac{\theta}{4\pi^2}\epsilon^{bi}\partial_i\lambda_b = 0
\end{equation}
Likewise, the dynamical piece of the action can be written as:
\begin{equation}
S_{dyn} = \frac{\theta}{4\pi^2}\int_\partial \frac{1}{2}\epsilon^{bi}(\dot{a}_{ij}a_b^{\,\,\,j} + \dot{\lambda}_i\lambda_b)
\end{equation}
It is noteworthy that, upon separating $a_{ij}$ into its trace $a^i_{\,\,i}$ and a traceless symmetric tensor $\tilde{a}_{ij}$, the trace component does not appear in either $S_{dyn}$ or the constraint equations, as can be readily checked.  The trace can therefore be eliminated from the theory entirely, leaving us with a decoupled theory of a vector field and a traceless symmetric tensor field.  The vector field $\lambda_i$ is governed by a standard Chern-Simons theory:
\begin{equation}
S_{dyn} = \frac{\theta}{8\pi^2}\int \epsilon^{bi}\dot{\lambda}_i\lambda_b
\end{equation}
\begin{equation}
\frac{\theta}{4\pi^2}\epsilon^{bi}\partial_i\lambda_b = 0
\end{equation}
The charges associated with this gauge field are simply the usual Chern-Simons charges, which obey charge conservation, but no other exotic conservation laws.

The traceless symmetric tensor $\tilde{a}_{ij}$ is governed by a chiral generalized Chern-Simons theory:
\begin{equation}
S_{dyn} = \frac{\theta}{8\pi^2}\int \epsilon^{bi}\dot{\tilde{a}}_{ij}\tilde{a}_b^{\,\,\,j}
\end{equation}
\begin{equation}
-\frac{\theta}{4\pi^2}\epsilon^{bi}\partial_i\partial_j\tilde{a}_b^{\,\,\,j} = 0
\end{equation}
We note that the two independent components of $\tilde{a}_{ij}$ are combined into one canonical conjugate pair.  Given that there is one constraint on $\tilde{a}_{ij}$, we see that the theory is fully constrained, and there are no local degrees of freedom.  We can also consider the appropriate gauge charges, $\epsilon^{bi}\partial_i\partial_j\tilde{a}_b^{\,\,\,j} = \rho$.  As can readily be checked, this charge has a conserved dipole moment.  We therefore have an example of a stable two-dimensional chiral fracton phase.  Bearing in mind the symmetry of $\tilde{a}^{ij}$, we can readily find the generalized Hall response to be:
\begin{equation}
\langle J^{ij}\rangle = \frac{\theta}{8\pi^2}(\epsilon^{ib}\dot{\tilde{a}}_b^{\,\,\,j} + \epsilon^{jb}\dot{\tilde{a}}_b^{\,\,\,i}) = \frac{\theta}{8\pi^2}(\epsilon^{ib}E_b^{\,\,\,j}+\epsilon^{jb}E_b^{\,\,\,i})
\end{equation}
If we define a generalized ``conductivity" tensor via $J^{ij} = \sigma^{ijk\ell}E_{k\ell}$, then this tensor will be given by:
\begin{equation}
\sigma^{ijk\ell} = \frac{\theta}{8\pi^2}(\epsilon^{ik}\delta^{j\ell}+\epsilon^{jk}\delta^{i\ell})
\end{equation}

We can verify the presence of an energy gap by regularizing the theory with a Maxwell term.  In this case, we include a Maxwell term appropriate to the \emph{traceless} scalar charge theory:
\begin{align}
\begin{split}
S = \int \bigg(\frac{1}{2e^2}((\dot{\tilde{a}}_{ij} - (\partial_i\partial_j - \frac{1}{3}\delta_{ij}\partial^2)\phi)^2 - (\epsilon^{ij}\partial_j\partial^k\tilde{a}_{ik})^2) \\
+ \frac{\theta}{8\pi^2}(\epsilon^{bi}\dot{\tilde{a}}_{ij}\tilde{a}_b^{\,\,\,j} - 2\phi\epsilon^{bi}\partial_i\partial_j\tilde{a}_b^{\,\,\,j})\bigg)
\end{split}
\end{align}
The corresponding equation of motion behaves as $\frac{1}{e^2}\dot{E}\propto \theta E$, giving us an energy gap scaling as $\Delta\sim \theta e^2$.  This Chern-Simons term is therefore capable of fully gapping the traceless scalar charge theory, yielding a stable two-dimensional phase of matter.  (Note that, alternatively, one might consider adding this same Chern-Simons term to the traceful scalar charge Maxwell theory, in which case the trace mode will remain gapless and the issue of stability is unclear.  The investigation of such a partially gapped theory is left as a task for the future.)

\subsection{Realization in Two Dimensions}

In this section, we have identified several tensor analogues of Chern-Simons theory, most of which are stable theories supporting either fractons or one-dimensional particles.  By looking at the boundaries of other higher-spin theories in three dimensions, we expect to discover many other generalized Chern-Simons theories, with tensors of arbitrary rank, which should generically host fracton excitations.

While we have discussed these theories as boundaries of three-dimensional higher-spin systems, we note that they can also occur in purely two-dimensional systems under appropriate conditions.  Most notably, certain values of $\theta$ correspond to a bulk which is identical to the $\theta = 0$ system ($e.g.$ when $\theta_1$ is a multiple of $2\pi$ in the vector charge theory).  At these specific values of $\theta$, the bulk is completely irrelevant to the boundary physics, and the generalized Chern-Simons theory is simply a decoupled theory sitting on top of the bulk.  This is closely analogous to Maxwell theory, where $\theta = 2\pi$ corresponds to a normal two-dimensional integer quantum hall layer deposited on the boundary.\cite{spt,chong}  Similarly, the boundaries of the higher-spin theories at these special values of $\theta$ host purely two-dimensional fracton theories, in a natural generalization of integer quantum hall states.  Indeed, some of these states can be intuitively understood as the mobile particles of the theory ($e.g.$ dipoles in the scalar charge theory) being put into a quantum Hall state of their emergent magnetic field.  This physical picture will be explored further in forthcoming work.

Values of $\theta$ other than these special ``integer" values naturally correspond to generalized fractional quantum hall systems.  In particular, the boundary theory will host excitations with a fraction of the fundamental charge of the bulk theory.  These fractional theories will also be able to occur in strictly two-dimensional systems, modulo issues of symmetry implementation.  In the story of conventional Chern-Simons theory, the only difference between a strictly 2d fractional quantum hall system and a boundary theory is in the implementation of symmetries (such as time reversal) on fractionalized particles, where issues concerning anomalies need to be carefully considered.\cite{spt,chong}  If the symmetry is allowed to be anomalous (for example, if an emergent symmetry like particle-hole symmetry is playing the role of time reversal), then these fractional boundary theories are perfectly valid in strictly two dimensions.  An entirely analogous story should hold in the case of the higher rank theories, but we leave the detailed consideration of symmetries and anomalies as a fight for future days.

The presence of fractons and other subdimensional particles in a purely two-dimensional system seems surprising at first.  For discrete fracton models, previous work has suggested that only three-dimensional (or higher-dimensional) models should be possible.\cite{fracton1,fracton2}  As for $U(1)$ fracton systems, theories involving only Maxwell-type terms suffer from instabilities to confinement in two dimensions.\cite{alex}  In the present case, we also have $U(1)$ fractons, but the two-dimensional gauge modes are gapped out via a Chern-Simons-like action.  This energy gap results in the stabilization of the theory, just as a Chern-Simons term stabilizes the conventional chiral $U(1)$ spin liquid in two dimensions.  We therefore see that we have a rather narrow path to obtaining fractons in two dimensions: the fractons must be $U(1)$ objects (integer-valued charge), and the corresponding $U(1)$ gauge theory must be stabilized against confinement by some mechanism.  The simplest means of stabilization is by adding Chern-Simons-like terms to gap the gauge field, as we have discussed.  We leave as an open question whether or not there are other valid 2d stabilization mechanisms, such as the addition of appropriate gapless matter fields.

\section{Conclusions}

In this paper, we have investigated the role of $\theta$ terms in the physics of three-dimensional higher-spin $U(1)$ gauge theories.  As in Maxwell theory, these higher-spin $\theta$ terms have two primary effects: the attachment of electric charge to magnetic monopoles (the Witten effect) and the presence of a nontrivial boundary theory.  We have worked out the details of the Witten effect for subdimensional particles, relating the $\theta$ parameter to the amount of electric charge attached to magnetic monopoles.  We have also studied the resulting boundary theory, which hosts fracton excitations coupled to a $U(1)$ gauge field, stabilized by generalized Chern-Simons terms.  These boundary theories thereby open the door to studying two-dimensional $U(1)$ fracton phases, both as boundary theories and in strictly two dimensions.

There remains interesting work to be done, particularly with the new two-dimensional phases.  We have identified some of the basic features characterizing these phases, such as their generalized Hall responses, but they surely possess features which still remain to be unlocked.  In particular, a purely two-dimensional chiral theory may host robust one-dimensional edge modes.  However, it remains unclear at present how precisely to study such edge modes.  It would also be interesting to study the role of global symmetries in the theories identified here, sorting out which symmetry implementations are possible, both with and without anomalies.  Such a study would be useful, not only for characterizing the new two-dimensional phases, but for characterizing the bulk behavior of the higher-spin $U(1)$ gauge theories in the presence of enrichment by global symmetries.  On the more down-to-earth side of things, it would be highly useful to find two-dimensional lattice models which demonstrate the properties of these phases explicitly, to complement the field-theoretic arguments presented here.  In addition, more investigation is required regarding how to measure the generalized Hall responses, which will be important for experimental detection of these two-dimensional fracton phases.

\section*{Acknowledgments}

I would like to acknowledge useful conversations with Kevin Slagle, Claudio Chamon, Rahul Nandkishore, Mike Hermele, Yizhuang You, Juven Wang, Abhinav Prem, Han Ma, T. Senthil, Liujun Zou, Yahui Zhang, and Yang-Zhi Chou.  This work was partially supported by a Simons Investigator Award to T. Senthil at the Massachusetts Institute of Technology and partially by the Center for Theory of Quantum Matter at the University of Colorado Boulder.

\section*{Appendix A: Mutual Statistics}

In the text, we claimed that, in the scalar charge theory, a magnetic vector charge and a parallel coplanar electric dipole have mutual $\pi$ statistics ($i.e.$ the wavefunction picks up a minus sign when one particle is wound around the other).  Assuming that the magnetic vector and the electric dipole are both bosons, the mutual $\pi$ statistics has the effect of making the bound state of the two into a fermionic object, as is familiar from the study of two-dimensional anyons.  We will now verify the claim of mutual $\pi$ statistics.

It is easiest to see the mutual statistics by considering winding the dipole around the magnetic vector (though winding the vector around the dipole should yield the same result).  As derived from previous work on tensor electromagnetism \cite{genem}, the effective magnetic field seen by a dipole $p^i$ takes the form $B_{eff}^i = -p_jB^{ij}$ (where the dipole moment is in units of the minimal dipole).  Now let us consider the magnetic field created by a magnetic vector charge $\partial_i B^{ij} = 2\pi g^j$.  The effective magnetic field seen by the dipole due to a magnetic vector charge then obeys:
\begin{equation}
\partial_iB_{eff}^i = -2\pi (p\cdot g)
\end{equation}
When the magnetic vector and electric dipole are parallel and are both unit strength, we see that $\partial_iB_{eff}^i = -2\pi$, meaning that the dipole sees the magnetic vector as a (negative) unit monopole of its effective magnetic field.  In other words, a surface enclosing the magnetic vector charge will see a $-2\pi$ flux of the effective magnetic field for the dipole.  The effective magnetic field corresponding to this monopole is not necessarily isotropic (it can depend on the orientations of $p^i$ and $g^i$), but when the dipole is wound around the vector in their plane of mutual motion, we can still conclude by symmetry that the enclosed flux (and the corresponding Aharonov-Bohm phase) is precisely half of the total, $-\pi$.  Thus, when an electric dipole is wound around a magnetic vector within their mutual plane of motion, the wavefunction of the system picks up a factor of $e^{-i\pi} = -1$, completing the proof of mutual $\pi$ statistics.

\section*{Appendix B: Are $\theta$ Terms Topological?}

In this work, we have considered natural generalizations of the $\theta$ term encountered in the study of Maxwell theory.  In the context of Maxwell theory, we can write this term in the form:
\begin{equation}
S_\theta = \int d^4x \epsilon^{\mu\nu\rho\sigma}F_{\mu\nu}F_{\rho\sigma} = \int F\wedge F
\end{equation}
where the last step takes advantage of the language of differential forms.  This rewriting makes it manifest that the $\theta$ term is topological, in that it does not rely on any special choice of coordinates or on the metric of the system.  This ensures that the properties of the $\theta$ term, such as being a total derivative and giving rise to the Witten effect, do not depend on whether we are considering a flat or curved spacetime.  Correspondingly, the boundary physics arising from the $\theta$ term is that of Chern-Simons theory, which is a topological quantum field theory.

In the main text, we have studied the generalized $\theta$ terms of the higher rank theories only at the level of flat space, with all indices raised and lowered by the flat metric, $\delta^{ij}$.  But given the formal similarity of these terms to the conventional $\theta$ term, it seems reasonable to ask whether or not the new $\theta$ terms also possess a topological character which would allow simple generalization to curved spaces.  Unfortunately, this does not appear to be the case.  First of all, unlike their antisymmetric cousins, symmetric tensors do not have a natural formulation in terms of differential forms, so there is no reason to expect any metric-independent formulation of the theory.  Furthermore, the boundary physics induced by the higher rank $\theta$ terms is that of tensor Chern-Simons theory, hosting fractons and other subdimensional particles.  Phases with such subdimensional particles tend to have large ground state degeneracies, growing with the system size\cite{fracton1,fracton2}.  This is clearly incompatible with the notion of a topological theory, where the physics should depend on the topology, but not the size, of the system on which it is defined.  Indeed, recent work\cite{slagle2} has indicated that fracton models have much greater sensitivity to the geometry of the system than a pure topological phase.  Fracton phases can even have robust degeneracy on a topologically trivial manifold, induced purely by the curvature of the system.

Similarly, the higher rank $\theta$ terms will likely feel the effects of geometry in subtle ways, making the extension to curved spaces a nontrivial one.  In this work, we will need to mostly content ourselves with the study of $\theta$ terms in approximately flat space, which should be sufficient for understanding the realization of these phases in the solid state context.  Nevertheless, the behavior of these theories in curved space is a question of both intrinsic interest and experimental relevance, since one can simulate the effects of curvature by putting the system under appropriate strains.  We leave the detailed investigation of higher rank phases in curved space to future work, though we can make some conjectures.  It is unclear if the higher rank $\theta$ terms remain total derivatives in curved space, or if they respond to curvature in such a way that changes the bulk physics of the gapless gauge mode.  It is entirely possible that the $\theta$ term will somehow modify the dispersion of the gauge mode, though we strongly suspect that it remains gapless.  We also expect the Witten effect to carry over unchanged, as this arises from a purely local constraint on the electric and magnetic fields, which should have no dependence on curvature.  The validity of these conjectures will need to be studied in detail in the future.

\section*{Appendix C: Gauge Fixing}

When calculating expectation values via the path integral formulation for a gauge theory, one unfortunate difficulty is that the ``plain vanilla" sort of path integral ($e.g.$ Equation \ref{plain}) integrates over all gauge-equivalent configurations of the gauge field, resulting in divergences which need to be tamed.  The simplest way to handle this problem is to introduce a gauge-fixing procedure explicitly into the path integral.  For standard abelian gauge theories, such as Maxwell theory, such gauge-fixing procedures are fairly mild, merely allowing for some freedom in the choice of the photon propagator.  For nonabelian gauge theories, however, additional complications arise, and one must account for extra ``ghost" fields in order to fully decouple the physical gluon from unphysical gauge degrees of freedom.  The higher rank gauge theories considered in this paper have been natural tensor analogues of abelian gauge theories, so one would guess that ghosts play no role in these theories.  We will here check this explicitly and verify that there is no need for ghost fields in the theories considered here.

\subsection*{Review of Maxwell Theory}

We begin by reviewing Maxwell theory, in which the bare path integral takes the form:
\begin{equation}
Z = \int \mathcal{D}A^\mu e^{i\int d^4x \frac{1}{4}F^{\mu\nu}F_{\mu\nu}}
\end{equation}
We can explicitly take gauge fixing into account by adding a resolution of the identity as follows:
\begin{equation}
Z = \int \mathcal{D}A^\mu\mathcal{D}\alpha \,\delta(G[A^\alpha])\det\bigg(\frac{\delta G[A^\alpha]}{\delta\alpha}\bigg) e^{i\int d^4x \frac{1}{4}F^{\mu\nu}F_{\mu\nu}}
\end{equation}
where $A_\mu^\alpha = A_\mu + \partial_\mu \alpha$ is a gauge-transformed field, and $G[A^\alpha] = \partial^\mu A_\mu^\alpha - \Lambda$ is a gauge-fixing condition, fixing the longitudinal component of $A$.  This extra determinant factor leads directly to ghost fields, which play a significant role in non-abelian gauge theories.  In an abelian gauge theory, however, we have the crucial fact that the determinant, $\det(\delta G/\delta\alpha) = \det(\partial^2)$, is a constant and can be brought outside the integral.  When calculating any expectation values, this constant will cancel out, indicating that ghost fields are completely decoupled from the physical gauge field.

\subsection*{A Higher Rank Example}

We now move on to consider a higher rank theory, specifically the scalar charge theory.  The bare path integral for this theory takes the form:
\begin{equation}
Z = \int \mathcal{D}A^{ij}\mathcal{D}\phi e^{i\int d^3xdt\mathcal{L}}
\end{equation}
where the Lagrangian was defined in Equation \ref{scalag}.  We can similarly add in a resolution of the identity as follows:
\begin{align}
Z = \int \mathcal{D}A^{ij}&\mathcal{D}\phi\mathcal{D}\alpha\,\delta(G[A^\alpha,\phi^\alpha]) \\
&\det\bigg(\frac{\delta G[A^\alpha,\phi^\alpha]}{\delta\alpha}\bigg) e^{i\int d^3xdt\mathcal{L}}
\end{align}
where $A^\alpha_{ij} = A_{ij} + \partial_i\partial_j\alpha$ and $\phi^\alpha = \phi + \dot{\alpha}$ are the gauge-transformed fields and $G = \partial_i\partial_j A^{ij} + \dot{\phi} - \Lambda$ is a gauge-fixing condition which fixes the unphysical components of $A$ and $\phi$.  Once again, we take advantage of the crucial fact that the determinant, $\det(\delta G/\delta\alpha) = \det((\partial_i\partial^i)^2 + (\partial_t)^2)$ is a constant and can be taken outside of the integral, therefore canceling out of any expectation values.  The same principle holds for any of the other higher rank theories, whose gauge transformations behave similarly.  We can therefore conclude that ghost fields decouple and have no role to play in any of the theories considered here, which justifies our neglect of them throughout this paper.


\begin{thebibliography}{9}
\bibitem{foot0}
The vector $\vec{g}$ must be an integer linear combination of the fundamental unit vector charges, which can depend on the underlying lattice.  On the cubic lattice, for example, $\vec{g}$ must take the form $(n_x,n_y,n_z)$ for integers $n_x$, $n_y$, and $n_z$.  For other lattices, different structures of charge lattice are possible.
\bibitem{foot1}
The other possibility, where $q$ is odd and $g$ is even, is simply related by duality.
\bibitem{foot2}
One could in principle also consider models where one or the other (or perhaps even both) of the magnetic vectors or electric dipoles are fermions.  The analysis in this case would be very similar, but likely has a few differences.
\end{thebibliography}
\end{document}